\documentstyle[12pt]{article}

\begin{document}


\title{Fractional Fokker--Planck Equation for  Nonlinear Stochastic
Differential Equations Driven by Non-Gaussian Levy Stable Noises}

\author{D. Schertzer,
M. Larchev\^eque \\
Laboratoire de Mod\'elisation en M\'ecanique, Tour 66, Boite 162
\\ Universit\'e
Pierre et Marie Curie,\\ 4 Place Jussieu, F-75252 Paris Cedex 05, France.\\
\and
J. Duan \\ Department of Applied Mathematics \\
Illinois Institute of Technology \\ Chicago, IL 60616 \\
\and
V.V. Yanovsky \\
Turbulence Research, Institute for Single Crystals\\
 National Acad. Sci. Ukraine, \\ Lenin ave. 60, Kharkov 310001, Ukraine\\
\and
S. Lovejoy \\ Physics Department, McGill University \\
3600 University Street
Montreal, H3A 2T8, Quebec, Canada.}

\date{ submitted to J. Math. Phys., October 28 1999 \
revised version June 29 2000}

\maketitle


\def\u{\underline}
\def\a{\alpha}\def\b{\beta}\def\g{\gamma}\def\d{\delta}\def\D{\Delta}
\def\e{\epsilon}\def\k{\kappa}\def\l{\lambda}\def\L{\Lambda}
\def\s{\sigma}\def\S{\Sigma}\def\Th{\Theta}\def\th{\theta}
\def\om{\omega}\def\Om{\Omega}\def\G{\Gamma}\def\v{\varepsilon}

\newtheorem{theorem}{Theorem}
\newtheorem{definition}{Definition}
\newtheorem{proposition}{Proposition}
\newtheorem{cor}{Corollary}
\newtheorem{lem}{Lemma}

\abstract{The Fokker-Planck equation has been very useful for studying
dynamic behavior of stochastic differential equations
driven by Gaussian noises. However, there are both theoretical
and empirical reasons to consider similar equations driven by
strongly non-Gaussian noises. In particular, they yield strongly
non-Gaussian anomalous diffusion which seems to be relevant in
different domains of Physics.
We therefore derive in this paper a Fractional Fokker--Planck equation
for the probability distribution of particles whose motion is governed by
a {\em nonlinear} Langevin-type equation,
which is driven by a Levy-stable noise rather than a Gaussian. We
obtain in fact a general result for a Markovian forcing.
We also discuss the existence and uniqueness of the solution of the Fractional
Fokker--Planck equation.}

\bigskip
\noindent Running title: Fractional Fokker--Planck Equation and
L\'evy noises\\

Correspondence should be addressed to D. Schertzer (fax: +33 1 44275259,
e-mail: schertze@ccr.jussieu.fr).
\newpage

\Roman{section}
\section{Introduction and motivation} \label{introduction}
The Fokker-Planck equation is one of the most celebrated equations in
Physics, since
it has been very useful for studying \cite{Kampen}
the dynamic behavior of stochastic differential equations
driven by Gaussian noises.
However, it turns out that many physical phenomena bring
this framework into question. For instance, it has been argued that diffusion
by geophysical turbulence \cite{Shlesinger,Solomon, Weeks, Ott, Duan,
Cherifi} corresponds,
loosely speaking, to a series of sticking (pauses), when the particle is
trapped
by a coherent structure, and (fast) flights, when the particle moves
in the jet flow.

Although there have been some attempts \cite{Duan} to analyze and
quantify this behavior with the help of the classical Fokker-Planck
equation, i.e. assuming finite moments of all orders,
some laboratory experiments \cite{Solomon, Weeks, Ott}
or numerical simulations of geostrophic turbulence \cite{del-Castillo}
show that this phenomenology could be rather a consequence of
the presence of heavy tails (i.e. power law fall-off) for the  probability
distribution and a strong
anisotropy with a clearly preferred direction of diffusion. One can
conclude \cite{Weeks.2} that if they are additive processes, the corresponding
walks are L\'evy motions.

Let us recall that indeed stable L\'evy motions $L(t)$ generalize the
Brownian motion $B(t)$
in the sense that first they are also motions (e.g. \cite{Fristedt,
Taquu}) whose increments
$\Delta L (t, \Delta t)= L(t + \Delta t)-  L(t)$ are
stationary (therefore $\Delta L$ has no statistical dependence on $t$) and
independent
for any non overlapping time lags $\Delta t$.  Therefore, $L(t)$
corresponds to the
sum of independent, identically distributed L\'evy stable variables
\cite{Levy, Khintchine, Gnedenko, Feller, Zolotarev}.
The second common property is that these increments satisfy a
``stability property'':  up to a rescaling and recentring, the sum of
different
steps has the same probability
distribution as one of the steps. L\'evy stable variables are precisely
defined by this property.
The stability property implies in both cases a
property of attraction: under rather general conditions a renormalized
sum of independent identically distributed variables converge to a
stable law. Furthermore, there are no other attractive laws.
This explains why the stable property is so important. The attraction
property corresponds to a broad generalization of the central limit
theorem, with the important difference that whereas the classical
theorem (Gaussian case) is satisfied with the condition that the variance
is finite,
the convergence towards a L\'evy law is obtained with the condition
that {\em not only} the variance of the summands $X_{i}$ is infinite, but
also that
all their moments of order $q$ equal to or larger than a critical order
$\alpha \: (0< \alpha <2)$
are infinite.  This critical order $\alpha$ is called
the L\'evy stability index and corresponds to the exponent of the power-law
of probability distribution tails:

\begin{equation}  \label{eq_pdf_tail}
any  \; s>> 1: \;  Pr(|\Delta L|>s) \approx s^{-\alpha} \; \leftrightarrow  \;
any  \;  q \geq \alpha:  E(|X|^q) = \infty
\end{equation}

\noindent where $Pr$ denotes the probability, $E(.)$ is the mathematical
expectation and $s$ is a given (large) nonnegative threshold.
This statistical divergence of a L\'evy motion is due to jumps,
whereas a Brownian motion is almost surely continuous.

This index is the most important of the four
parameters defining a L\'evy stable law. The second one is the
`skewness' $\beta \:( -1 \leq
\beta \leq  1)$ which defines the degree of asymmetry of the law, which is
maximal for $\beta = - 1$ or $\beta = + 1$, and the law is symmetric
when  $\beta = 0$. In spite of its name and some common properties,  $\beta $
nevertheless does not correspond to the classical skewness of a
quasi-Gaussian law.
The latter is indeed undefined for a stable L\'evy law due to the above
mentioned statistical divergences.
The center $\gamma $ corresponds to the statistical mean when
defined (i.e. $\alpha > 1$) and/or to the median when symmetric
(i.e. $\beta = 0$). The scale parameter $D$ ($D \geq 0$) corresponds
to a generalization of the variance of the Gaussian case. More
precisely, as discussed below, it corresponds to  the intensity
scale of the cumulant of (possibly non integer) order $\alpha$.
It yields an anomalous \cite{Yan} generalization of the  classical Einstein
relation:
$ Var[X(t) -X(t_0)] = 2 D (t-t_0)$, where $ Var(.)$ denotes the variance.
Finally, let us emphasize that
the Gaussian case corresponds to the limit case $\alpha =2$, which also
implies $\beta = 0 $, i.e. no asymmetry.

Further comments are now in order on the relevance of L\'evy
motions in Physics. On the one hand, claims in favor of the
relevance of L\'evy motions have been made on many physical phenomena
ranging from subrecoil
laser cooling \cite{Bardou, Reichel} to diffusion by flows in porous
media \cite{Painter, Benson}, including finance fluctuations
\cite{Stanley,Schmitt}, see Ref.\cite{Shlesinger2, Shlesinger3} for other
examples.
 Many systems indeed display  a phenomenology rather similar to
that we reported above on geostrophic turbulence.

On the other hand, important questions have been raised.  In particular,
Ref.\cite{Boland}
questioned the resulting infinite variance of the advecting
field for porous media. Indeed, it turns out that recent
estimates \cite{Tchiguirinskaia} of the power-law of the probability
distributions of the
hydraulic conductivity yields an exponent  $\alpha \approx 3.5$.
The question of finite variance might apply to other examples, in
particular for atmospheric
turbulence where different studies \cite{turbulence} yield  a critical
exponent
$\alpha \approx 7$ for the wind field.  Therefore, in spite of their
clear phenomenological interest,
the relevance of pure L\'evy motions could be questioned.

The main goal of this paper is to clarify and define a
framework adequate  for handling motions more general than pure L\'evy
motions and which are nevertheless generated by the latter.
We will do it by building upon a series of rather
recent works
\cite{Zaslavsky, Fogedby, Chechkin, Compte, Marsan, Chaves, Yan, Meerschaert}
which show that the probability
density of particles moving with a Levy motion satisfies a generalized
Fokker-Planck
equation involving fractional orders of differentiation.
Indeed, it could be first
argued in a ``very formal and phenomenological'' manner \cite{Zaslavsky}
that a fractional power of the Laplacian
yields an anomalous scaling for the corresponding diffusion.

A Fractional Fokker-Planck equation
was obtained in a less formal manner by \cite{Fogedby,Compte} in the
framework of
the continuous time random walks (CTRW's) model of anomalous diffusion
\cite{Klafter}.
However, this method does not involve directly a stable L\'evy process, but
a walk
sharing some behavior common with the latter, without being equivalent to it.
A different Fractional Fokker-Planck equation was introduced \cite{Chaves}
with the help of a phenomenological and interesting transformation of the
classical Fick law
into a fractional Fick law. However, it is not clear that its solution
corresponds to a (nonnegative) probability distribution.  A rather distinct
approach was followed by  \cite{Chechkin, Yan}
since it starts with a {\em linear} Langevin-like
equation with random forces which are  {\em exact} stable Levy processes,
which can be symmetric as well as asymmetric, and with no limitation on
the possible values of the Levy index $\alpha$. The fundamental
mathematical tool which is used is the second characteristic (or
cumulant generating)
function of the motion defined by this Langevin-like equation.
The particular case of symmetric processes
correspond to what was previously inferred by
\cite{Zaslavsky,Fogedby,Compte,Chaves}.
However, it was shown that in the more general case of asymmetric  processes,
a new non-trivial `advective-diffusive' term appears. This is
confirmed with the help of a re-interpretation of the characteristic
function of a Levy
motion \cite{Meerschaert}.

We already discussed that theoretically and empirically the non finiteness
of the variance
could be questioned. There are two more general questions:
the inhomogeneities of the medium, which are first emphasized for the
introduction of the L\'evy motions, are finally reduced to a
(homogeneous) distribution of times when the particle is strongly
kicked. As soon as this representation is granted, the medium (and its
properties)
does not intervene any longer.
This is very restrictive and for instance incompatible with the
multifractality
of the medium (or of the diffusion) when observed. The second
reason is that the underlying processes are thought to be strongly
nonlinear, whereas the transport is modeled with the help of a
(stochastic) linear equation.

Both the successes and limitations of the previous results plead in
favor of investigating a local and nonlinear modeling with the help of L\'evy
motions. This is the reason that we investigate the properties a of {\em
nonlinear}
Langevin--like equation forced by a L\'evy stable motion.

\section{Statement of the problem} \label{statement}

Further to our above discussion, we consider the following
{\em nonlinear} Langevin--like equation for a
stochastic (real) quantity $X(t)$ (e.g. location of a particle):

\begin{equation}  \label{eq_langevin}
d X(t) = m(X(t),t)~ dt + \sigma(X(t),t)~ dL
\end{equation}

\noindent where the driving source is a L\'evy stable motion $L(t)$
instead of Brownian motion  $B(t)$. The latter case corresponds
to the basis of stochastic calculus (e.g. Ref.\cite{Doob}) and the
corresponding differential equation is often called the Ito-Skorokhod
equation.The extension
to L\'evy stable motion $L(t)$ is rather natural and
straightforward (e.g. \cite{Janicki}) due to the common properties of
$L(t)$ and  $B(t)$ that we discussed in Sect. \ref{introduction},
i.e. their infinitesimal increments are independent identically distributed
and
furthermore stable.

More precisely the Ito stochastic calculus corresponds to consider
that the  $d L$ is, similarly to $d B$, a forward increment in time (it should
be understood as $d L(t,dt) = L(t+ dt)-L(t)$). This means
that the value of $X$ at time $t$ is determined by events prior to the
application of
the stochastic force  $d L(t)$, which acts only from time $t$ to $t + dt$.

The Eq. \ref{eq_langevin} can also be understood under its integral form:

\begin{equation}  \label{eq_integral}
X(t) = X(t_{0}) +\int m(X(t),t) dt + \int \sigma(X(t),t)~ dL
\end{equation}

\noindent where the last term corresponds to a stochastic integration of a
stochastic process.
The integration of a stochastic process
$\Phi (t)$ (in the case of  Eq. \ref{eq_langevin}: $\Phi (t) =
\sigma(X(t),t))$ by the L\'evy motion $L$, is rather straightforward
in the case of step processes \cite{Kwapien}:

\begin{eqnarray}  \label{simple_processes}
 \Phi (t) =\Phi_{n}, for \; t \in  (t_{n}, t_{n+1}), n=0,1,\ldots N-1
 \nonumber \\
\int \Phi (t)~ dL = \sum_{n=0}^{N-1} \Phi_{n} (L(t_{n+1})- L(t_{n}))
\end{eqnarray}

\noindent and this rather suggestive definition is naturally extended to
functional spaces in
which the step processes are dense.

In order to establish local
properties, for instance the time evolution of the probability
of the particles, we will use the
differential form (Eq. \ref{eq_langevin}), whereas \cite{Chechkin, Yan}
rather used the integral form (Eq.\ref{simple_processes}) which
becomes cumbersome in the nonlinear case and is in fact useful only
to establish global properties (Sect.\ref{sect.existence}).

After having emphasized the similarities between $L(t)$ and $B(t)$, it is
important to underline
the non-trivial consequences due to the fact,
contrary to the Gaussian case which has all its moments finite,
L\'evy motions have a finite critical order of divergence of statistical
moments
($0<\alpha <2$). These include the fact that the mathematical
techniques which could be used can be rather distinct.
For instance, our derivation will rely on the
use of the second characteristic function of the increments
Sect.\ref{cumulants.sect}, instead of
probabilities of the increments as done usually for the derivation of
the classical Fokker-Planck equation. An obvious reason is that the
former are relatively simple (see Sect.\ref{Levy case}), while the latter are
not, with the only exception of the three following cases:
$\a =2$, $ \b =0$; $\a =1, \b =0$; $\a =1/2, \b =1$. The fundamental
reason is that both the stability property and the divergence of moments
are related to the presence of a cumulant of non integer order $\alpha$.
In relation to this problem, the convenient $L^{2}$ Hilbert structure
of Gaussian processes
is reduced to a $L^{\alpha}$ Banach structure for stable L\'evy
processes. This is particularly important for the
integral equation Eq.\ref {eq_integral}, when defining functional
spaces where step processes are dense.

The linear case, which is the hitherto studied case, corresponds to:

\begin{equation}  \label{eq_linear_case}
m(X(t),t) \equiv m = Const., ~~ \sigma(X(t),t) \equiv \sigma = Const.
\end{equation}

\noindent $X(t) -X(t_0)$ is also a Levy motion which has the same Levy
stability
index $\alpha $ as its increments,
but with a different center or trend and scale or amplitude.

In the nonlinear case,
 $m(X(t),t)$ and $\sigma (X(t),t)$ are (possibly nonlinear)
functions of $X(t)$ and $t$, which satisfy certain regularity constraints to
be discussed later (Sect.\ref{sect.existence}). They correspond to
inhomogeneities of the medium, which were ignored in the linear case.
As a possibly important, but simple example,
let us mention the L\'evy extension of the so-called geometric Brownian
motion,
which is rather ubiquitous and for instance is at the core of the
Black-Scholes model for option pricing:
$m(X(t),t)= m X(t)$ and $\sigma (X(t),t)= \sigma X(t)$, where
 $\sigma$ is the `votality' constant of the price $X(t)$ of a given stock
share.

We will demonstrate that:

\begin{proposition}\label{prop.1}
\begin{quote} The transition probability density:

\begin{equation} \label{eq.trans_proba}
\forall t\geq t_{0}:~~p(x,t|x_{0},t_{0})=Pr(X(t)=x|X(t_{0})=x_{0})
\end{equation}

\noindent corresponding to the nonlinear stochastic differential equation
(Eq.\ref{eq_langevin}), with $\alpha \neq 1 ~or~ \beta = 0$,
 is solution of the
following Fractional Fokker-Planck equation: .

\begin{eqnarray}  \label{eq_fokker_planck}
{\frac{\partial }{\partial t}} p(x,t|x_0,t_0) = - {\frac{{\partial} }{{%
\partial x}}}(\gamma \sigma (x,t) + m(x,t)) p(x,t|x_0,t_0)   \nonumber \\
 -  D [(-\Delta )^{\alpha /2}  (|\sigma (x,t)|^{\alpha}
 p(x,t|x_0,t_0) )   \nonumber \\
 + \beta \omega (\alpha ) {\frac{\partial }{\partial x}} (-\Delta )^{(\alpha
-1)/2} (|\sigma (x,t)|^{\alpha -1} \sigma (x,t) p(x,t|x_0,t_0))]
\end{eqnarray}
\noindent with the initial condition:

\begin{equation} \label{eq.initial}
p(x,t_0|x_0,t_0) = \delta(x-x_0)
\end{equation}

\noindent where $\delta(x-x_0)$ is the degenerate Dirac measure in $x_0$ and
$\omega(\alpha)$ is defined by:

\begin{equation}  \label{eq_omega}
\alpha \neq1:~~ \omega (\alpha ) = tan {\frac{\pi \alpha }{2}}
\end{equation}

\end{quote}
\end{proposition}

\noindent  and where the fractional powers of the Laplacian $\Delta$ will
be discussed
in Sect.\ref{sec.fractional}. Proposition \ref{prop.1} and
Eq.\ref{eq_fokker_planck}
are for scalar processes
(i.e. $\Delta \equiv  {\frac{\partial^2 }{\partial x^2}}$)
and their extension to vector processes will be discussed and presented in
Sect.\ref{sec.high.dim}.
One may note that the fractional diffusive isotropic operator $-(-\Delta
)^{\alpha /2}$ applies via
a fractional diffusivity $|\sigma (x,t)|^{\alpha}$, whereas the
advective-diffusive term corresponds to a conjugate action of a
fractional diffusive term $-(-\Delta )^{(\alpha-1)/2} |\sigma (x,t)|^{\alpha
-1}$ and a convective term ${\frac{\partial }{\partial x}}\sigma (x,t) $
on the transition probability.

This Fractional Fokker-Planck equation
will be established with the help of the much more general proposition:

\begin{proposition}\label{prop.2}
\begin{quote}
The inverse Fourier transform of the second characteristic
function or cumulant generating function of the increments of a Markov
process $X(t)$ generates by convolution the Fokker-Planck equation of
evolution of its
transition probability $p(x,t|x_{0},t_{0})$.
\end{quote}
\end{proposition}

We will demonstrate this proposition in a straightforward, yet rigorous way.
More precisely, we will establish the
following:

\begin{equation} \label{eq.FP.generation}
{\frac{\partial p}{\partial t}}(x,t|x_{0},t_{0}) =
\int {dy {\frac{\partial \widetilde{K}}{\partial t}}(x-y|y,t)
p(y,t|x_{0},t_{0})}
\end{equation}
\noindent where $\widetilde{K}$ is the inverse Fourier transform of the
cumulant generating
function of the increments. The  $\widetilde{K}$ arguments will become
explicit in Sect.\ref{cumulants.sect}.

Eq.\ref{eq.FP.generation} not only holds for
processes with stationary and independent increments,
as in the linear case (Eq.\ref{eq_linear_case})
but also for any Markov process, including those defined by
the non-linear Langevin-like equation (Eq.\ref{eq_langevin}
with $m \ne Const.$, $\sigma \ne Const.$).
As a consequence of Eq.\ref{eq.FP.generation}, we will demonstrate the
following:

\begin{proposition}\label{prop.3}
\begin{quote}
When the increment's cumulant generating function of a Markov process
$X(t)$ is defined by its expansion in cumulants $C_{n}$, its
Fokker-Planck equation is :
\end{quote}

\begin{equation}\label{eq_NFP_markov.3}
{\frac{\partial p}{\partial t}}(x,t|x_{0},t_{0})=\sum_{n \in J}
 {\frac{(-1)^n} {n!}} {\frac{\partial ^{n}}{\partial x^{n}}}
[ C_{n}(x,t) p(x,t|x_{0},t_{0})]
\end{equation}
\end{proposition}

An obviously sufficient condition of convergence is obtained when
the set $J$ of the orders  of differentiation $n$ is finite. This is
true in particular for Gaussian forcing: $J = \{1, 2\}$.
It corresponds to the classical Fokker-Planck equation.
On the other hand, $J ={\bf N }$ would correspond to an analytic expansion of
cumulants. In spite of its interest,
we will not discuss the latter case in this paper, nor its relationship to
the classical
Kramers-Moyal expansion  (e.g. \cite{Gardiner}).

Below, we concentrate on the case of
a finite, but non analytic expansion:  $J =\{1, \alpha \}$
(non integer $\alpha, \: 0< \alpha < 2$), since it corresponds to the L\'evy
extension (Sect.\ref{Levy case} and yields Prop.\ref{prop.1}
with the help of fractional derivatives, as discussed in
Sect.\ref{sec.fractional}.

\section{The cumulant generating function of the increments}
\label{cumulants.sect}

The first and second (conditional) characteristic functions are
respectively the moment generating function $Z_{X}(k,t-t_{0}|x_{0},t_{0})$ and
the cumulant generating function $K_{X}(k,t-t_{0}|x_{0},t_{0})$, associated
with the transition probability $p(x,t|x_{0},t_{0})$ of a process $X(t)$.
These are defined
by the Fourier transform of the latter, with $k$ being the conjugate
variable of $x-x_{0}$ :

\begin{eqnarray}  \label{eq_fc_Z}
F[p(x,t|x_{0},t_{0})] &\equiv & Z_{X}(k,t-t_{0}|x_{0},t_{0})  \\
&\equiv & exp(K_{X}(k,t-t_{0}|x_{0},t_{0}))   \\
&\equiv & E [exp(ik(X(t)-X(t_{0}))| X(t_{0}) = x_{0}]
\end{eqnarray}

\noindent where  $E[ \cdot | \cdot ]$ denote the conditional mathematical
expectation,
$F$ and $F^{-1}$  respectively the Fourier--transform
and its inverse:

\begin{eqnarray}
F[f] &=&\hat{f}(k)=\int_{-\infty }^{\infty }{dx}~~exp(ikx)f(x)~~~
\label{eq.fourier} \\
~F^{-1}[\hat{f}] &=&f(x)=\int_{-\infty }^{\infty }{\frac{dk}{2\pi }}%
~~exp(-ikx)\hat{f}(k)
\end{eqnarray}

The corresponding quantities for increments $\delta X(\delta t)=X(t +
\delta t)-X(t)$,
corresponding to a given time lag $\delta t>0$, are defined in a
similar way:

\begin{eqnarray}
F[p(x +\delta x , t+\delta t|x,t)] &=&\delta Z_{X}(k,\delta t|x,t)
\label{eq_incr} \\
&\equiv & exp(\delta K_{X}(k,\delta t|x,t)) \\
& = & E[ exp(ik (X(t + \delta t)- X(t))|  X(t)= x]
\end{eqnarray}

\noindent where $k$ is the conjugate variable of $\delta x$.
The cumulants of the increments $C_{n}$
are the coefficients of the
Taylor expansion of ${\delta K_{X}}$:

\begin{equation} \label{eq.cumulant_gen}
{\delta K_{X} (k,\delta t|x,t)=\delta t}
\sum_{n \in J}{\frac {(ik)^{n}} {n!}}C_{n}(x,t)+o({\delta t)}
\end{equation}

As already mentioned, the classical case corresponds to
an analytic expansion of ${\delta K_{X}}$, i.e. $J \subseteq {\bf N}$, whereas
we will be interested by a finite but non-analytic expansion $J= \{1, \alpha
\}$ (non integer $\alpha, \: 0< \alpha < 2$).

\section{Processes with stationary and independent increments}
\label{sect.sii_porcesses}
Let us first consider the simple sub-case of a process with
stationary and independent increments. It corresponds to
$ C_{n}(x,t) \equiv  C_{n} = Const.$ in Eqs.\ref{eq_NFP_markov.3},
\ref{eq.cumulant_gen} and as already discussed in Sect. \ref{introduction},
it includes the linear case (Eq.\ref{eq_linear_case}) of the Langevin--like
equation (Eq.\ref{eq_langevin}).

However, we believe that the following derivation is not
only somewhat pedagogical on the role of the characteristic functions
for the nonlinear case, but also terser than
derivations previously presented for the linear case.

The stationarity of the increments implies that the transition probability
depends only on the time and
space lags, i.e.:

\begin{equation} \label{eq.pdf_ind_incr}
p(x,t|x_{0},t_{0})=p(x-x_{0},t-t_{0})
\end{equation}

\noindent and similarly, the characteristic
functions of the increments are no longer conditioned, for instance:

\begin{eqnarray}
Z_{X}(k,t-t_{0}|x_{0},t_{0}) &\equiv & Z_{X}(k,t-t_{0})  \\
K_{X}(k,t-t_{0}|x_{0},t_{0}) &\equiv & K_{X}(k,t-t_{0})
\end{eqnarray}

On the other hand, the independence of the increments
implies that the transition probabilities satisfy a convolution (over any
possible intermediate
position $y$) for any  given time lag $\delta t$:

\begin{equation} \label{eq.convolution}
\forall  \delta t >0 : p(x-x_{0},t+ \delta t-t_{0}) = \int dy ~~
p(x-y,\delta t)p(y - x_{0},t-t_{0})
\end{equation}

\noindent and the corresponding characteristic functions merely
factor (resp. add). Therefore, we have:

\begin{equation} \label{eq.fc_ind_incr}
Z_{X}(k,t+\delta t-t_{0})-Z_{X}(k,t-t_{0})=
Z_{X}(k,t-t_{0})({\delta Z_{X}(k,\delta t)-1)}
\end{equation}

This in turn leads to:

\begin{equation} \label{eq.fourier_chap_kolmo}
Z_{X}(k,t+\delta t-t_{0})-Z_{X}(k,t-t_{0})=
Z_{X}(k,t-t_{0}) {\delta K_{X}(k,\delta t)} + o(\delta t)
\end{equation}

Its inverse Fourier transform yields:

\begin{equation} \label{eq_NFP_ind.inc.1}
p(x,t+\delta t|x_{0},t_{0})-p(x,t|x_{0},t_{0}) =
\int {d y} F^{-1}[\delta K_{X}(k,\delta t)]
p(y-x_{0},t-t_{0}) + o(\delta t)
\end{equation}

This demonstrates (in the limit $\delta t\to 0$)
Prop.\ref{prop.2} and Eq.\ref{eq.FP.generation},
as well as Prop.\ref{prop.3}, since Eq.\ref{eq_NFP_ind.inc.1} corresponds,
with the help of Eq.\ref{eq.cumulant_gen}, to:

\begin{equation} \label{eq_NFP_ind.inc.2}
p(x,t+\delta t|x_{0},t_{0})-p(x,t|x_{0},t_{0}) =\delta t\sum_{n \in J}
{\frac {(-1)^{n}} {n!}} [C_{n}
\int dy\delta _{x-y}^{(n)} p(y,t|x_{0},t_{0})] + o(\delta t)
\end{equation}

\noindent  where $\delta _{x}^{n}$ denotes
the $n^{th}$ derivative of the Dirac function. Therefore, we obtain:

\begin{equation} \label{eq_NFP_ind.inc.3}
{\frac {\partial }{\partial t}} p(x,t|x_0,t_0) = \sum_{n \in J}
{\frac {(-1)^{n}} {n!}} C_{n} {\frac{\partial ^{n}}{\partial x^{n}}}
p(x,t|x_{0},t_{0})
\end{equation}

\noindent which corresponds to the linear case of Eq.\ref{eq_NFP_markov.3}.

\section{More general Markov processes} \label{sect.Markov}

In the case of a  Markov process which does not have stationary and
independent increments,
there is no longer a simple convolution equation (Eq. \ref{eq.convolution}) of
the transition probabilities, nor
a simple factorization of characteristic functions
(Eq.\ref{eq.fc_ind_incr}).
However, the former satisfies a generalized convolution equation which
corresponds to the
Chapman-Kolmorogorov identity \cite{Feller} valid
for any Markov process $X(t)$:

\begin{equation} \label{eq.chap_kolmo}
\forall \delta t > 0:  p(x,t+\delta t|x_{0},t_{0}))=\int {d y
~p(x,t+\delta t|y,t)p(y,t|x_{0},t_{0})}
\end{equation}

\noindent which indeed reduces to a mere convolution  (Eq.
\ref{eq.convolution})
in the case of processes with
stationary and independent increments. This identity can be written under
the equivalent form:

\begin{equation}
p(x,t+\delta t|x_0,t_0) = \int{d y \int{{\frac{dk }{2\pi }} e^{-i k
y + \delta K_{X}(k,\delta t |y,t )}} p(y,t|x_0,t_0)}
\end{equation}

Noting that we have:

\begin{equation}
p(x,t|x_{0},t_{0}) =\int {dy~p(y,t|x_{0},t_{0})}%
\int {{\frac{dk}{2\pi }}e^{-ik y}}
\end{equation}

\noindent we obtain:

\begin{equation} \label{eq_NFP_markov.1}
p(x,t+\delta t|x_{0},t_{0})-p(x,t|x_{0},t_{0}) = \delta t \int {d y}
F^{-1}[\delta K_{X}(k,\delta t|y,t)]
p(y,t|x_{0},t_{0}) + o(\delta t)
\end{equation}

In the limit $\delta t\to 0$, this corresponds to
Prop. \ref{prop.2} and Eq. \ref{eq.FP.generation}.
When $J \subseteq {\bf N}$, it yields with the help of
Eq.\ref{eq.cumulant_gen}:

\begin{equation} \label{eq_NFP_markov.2}
\delta p(x,t|x_{0},t_{0})=\delta t\sum_{n \in J} \int {d y} \delta
_{x-y}^{(n)}
[{\frac {(-1)^{n}} {n!}}C_{n}(y,t)p(y,t|x_{0},t_{0})]+o(\delta t)
\end{equation}

The limit $\delta t \to 0$
corresponds to Eq.\ref{eq_NFP_markov.3} and demonstrates Prop.
\ref{prop.3} for a Markow process.

\section{Extension to fractional orders}\label{sec.fractional}

In the two previous sections (Sects.\ref{sect.sii_porcesses}-
\ref{sect.Markov}),
the fact that the indices $n \in J$ should be integers intervene at
best only in the correspondence between (integer order) differentiation
${\frac {\partial ^{n}}{\partial x^{n}}}$ (in Eq. \ref{eq_NFP_markov.3})
and powers of the conjugate variable $k^n$ (in Eq. \ref{eq.cumulant_gen}).
However, by the very definition
of fractional differentiation (e.g.\cite{Miller}), this correspondence holds
also for non integer orders. However, there is not a unique definition of
fractional differentiation and therefore, as discussed in some details in
\cite{Yan}),
we cannot expect to have a unique expression of the Fractional Fokker-Planck
equation.

Since in the following it will be sufficient to consider
an expansion of the characteristic function involving fractional powers of
only the wave number $|k|$, it is interesting to  consider
Riesz's definition of a fractional differentiation.
Indeed, the latter corresponds to consider
fractional powers of the Laplacian:

\begin{equation}\label{eq_riesz}
-(-\Delta )^{\alpha /2}f(x)=F^{-1}[|k|^{\alpha}\hat{f}(k)]
\end{equation}

\noindent which has furthermore the advantage of being valid for the vector
cases.
However, we will see in Sect. \ref{sec.high.dim} that in general it does
not apply in a
straightforward manner for vector stable L\'evy motions. Indeed the latter
introduces
rather (one-dimensional) directional Laplacians, i.e. (one-dimensional)
Laplacians along a
given direction $\u{u}$ ($\mid \u{u} \mid =1$):

\begin{equation} \label{eq_direct_laplacian}
-(-\Delta_{\u{u}})^{\alpha /2}f(x)=F^{-1}[|(\u{k},\u{u})|^{\alpha}\hat{f}(k)]
\end{equation}

\noindent where(.,.) denotes the scalar product. On the other hand, it will
be useful
to consider the fractional power of the contraction of the Laplacian tensor
$\u{\u{\Delta}}$:

\begin{equation} \label{eq_laplace_tensor}
\Delta_{i,j} = {\frac{\partial }{\partial x_i}} {\frac{\partial }{\partial
x_j}}
\end{equation}

\noindent by a tensor $\u{\u{\sigma }}$ ($\u{\u{\sigma}}^{*}$ denotes
its transpose), with the following definition:

\begin{equation} \label{eq_sigma_laplacian}
-(-\u{\u{\Delta}}:\u{\u{\sigma}}.\u{\u{\sigma}}^{*})^{\alpha \over 2}
\equiv F^{-1} [\mid (\u{k}, \u{\u{\sigma }}.\u{\u{\sigma }}^{*}
.\u{k}\mid^{\alpha \over 2}]
=  F^{-1} [\mid \u{\u{\sigma }}^{*} .\u{k}\mid^{\alpha}]
\end{equation}

\section{Levy case}\label{Levy case}

The second characteristic function  of the increments $\delta L$
of the (scalar) Levy forcing is the following:

\begin{equation}  \label{eq_fc_dL}
\delta K_{L}(k,\delta t) =
\delta t [ik\gamma - D|k|^\a ( 1- i\beta {\frac{k}{|k|}} \omega(k, \alpha)
] + o(\delta t)
\end{equation}

\noindent where $\omega(k, \alpha)$ is defined by:

\begin{equation}  \label{eq_omega_k}
\alpha \neq1:~~ \omega (k,\alpha ) \equiv \omega (\alpha ) = tan {\frac{
\pi \alpha }{2}}; ~~~~\alpha = 1:~~\omega (k,\alpha ) ={\frac{\pi }{2}}
log|k|
\end{equation}

Considering an Ito-like forward integration of Eq.\ref{eq_langevin}, the
increments $\delta L$
generates the following (first) characteristic function for the increments
$\delta X$ of the motion
$X(t)$:

\begin{equation} \label{eq_fc_dZ}
\delta Z_{X}(k,\delta t|x- \delta x, t)  = E(e^{ik m(X,t)})
\delta Z_{\sigma L}(k,\delta t|x, t) + + o(\delta t)
\end{equation}

\noindent which yields the following elementary cumulant generating
function $\delta K_{X}$:

\begin{eqnarray}  \label{eq_fc_dK}
\delta K_{X}(k,\delta t|x,t)=
{\delta t}
 [ik m(x,t) + ik\gamma \sigma (x,t)  \nonumber \\
 - D|k|^\a |\sigma (x,t)|^{\alpha}(1- i\beta {\frac{k \sigma (x,t)}{|k||\sigma (x,t)|}} \omega(k, \alpha)) ]
+ o(\delta t)
\end{eqnarray}

\noindent and which is of the same type as Eq.\ref{eq.cumulant_gen}, with $J=
\{1,\alpha \}$. Therefore, as discussed in Sect.\ref{sec.fractional}, we
have fractional differentiations in the corresponding
Eq.\ref{eq_NFP_markov.3},
which will precisely correspond to Eq.\ref{eq_fokker_planck}, and therefore
establishes Prop. \ref{prop.1}.

\section{Extension to vector processes}\label{sec.high.dim}

With but one important exception, the extension of the previous
results to higher dimensions is rather straightforward. The starting
point of this extension is the following  nonlinear stochastic equation
($\u{X}(t) \in R^d$):

\begin{equation}  \label{eq_ddim_langevin}
d \u{X}(t) = \u{m}(\u{X}(t),t) dt + \u{\u{\sigma}}(\u{X}(t),t).d\u{L}
\end{equation}

\noindent where $\u{m}$ is the natural d-dimensional vector extension
of the deterministic-like trend, $\u{\u{\sigma}}$ is the dxd'-dimensional
tensor extension of the modulation of the random driving force, and
$\u{L}$ is a d'-dimensional Levy stable motion. As discussed below,
the expression of the characteristic function of the latter
corresponds to the source of the difficulty in extending the scalar results
to higher dimensions.
On the contrary, it is straightforward to check that
Props. \ref{prop.2}, \ref{prop.3} are valid in the vector case,
with the following extensions ($\u{x} \in R^d$)  for Eq.
\ref{eq.FP.generation}:

\begin{equation} \label{eq.ddim.FP.generation}
{\frac{\partial p}{\partial t}}(\u{x},t|x_{0},t_{0}) =
\int {dy {\frac{\partial \widetilde{K}}{\partial t}}(\u{x}-\u{y}|\u{y},t)
p(\u{y},t|x_{0},t_{0})}
\end{equation}

\noindent and for Eq. \ref{eq_NFP_markov.3} ($\u{n}\in J \subseteq {\bf
N}^d, \mid \u{n} \mid = \sum_{i=1}^{d} n_i $):

\begin{equation} \label{eq_ddim_NFP_markov.3}
{\frac{\partial p}{\partial t}}(\u{x},t|\u{x}_{0},t_{0})=\sum_{\u{n} \in J}
  {\frac{(-1)^{\mid \u{n} \mid}} {(n_1)!(n_2)!..(n_d)!}}{\frac {\partial
^{\mid \u{n} \mid}}
  {\partial x_{1}^{n_1} \partial x_{2}^{n_2}..\partial x_{d}^{n_d}}}
[C_{\u{n}}(\u{x},t) p(\u{x},t|\u{x}_{0},t_{0})]
\end{equation}

On the other hand, Eq. \ref{eq_ddim_langevin} yields the following
extension to Eq.\ref{eq_fc_dZ}:

\begin{equation} \label{eq_ddim_fc_dZ}
\delta Z_{\u{X}}(\u{k},\delta t|\u{x}, t)  = e^{i~\u{k}.\u{m}(\u{x},t)}
\delta Z_{\u{\u{\sigma}}.\u{L}}(\u{k},\delta t|\u{x}, t)
\end{equation}

\noindent and therefore we have:

\begin{equation} \label{eq_ddim_fc_dK}
\delta K_{\u{X}}(\u{k},\delta t|\u{x}, t)  = i~\u{k}.\u{m}(\u{x},t) +
\delta K_{\u{L}}(\u{\u{\sigma}}^{*}.\u{k},\delta t|\u{x}, t) + o(\delta t)
\end{equation}

Let us recall that a stable L\'evy vector in the classical sense
\cite{Levy, Paulauskas, Nikias} (see \cite{Generalized_levy_vector}
for a discussion on a rather straightforward generalization, or
\cite{Siegel,Meerschaert2,Jurek}
for a more abstract generalization) corresponds to the limit of a sum of
jumps,
with a power-law distribution, along
random directions $\u{u} \in \partial B_1$, $B_1$ being the unit ball,
distributed according to a (positive) measure $d\Sigma(\underline{u})$.
The latter, which generalizes the scale parameter $D$ of the scalar case,
is the source of the difficulty since in general the probability
distribution of a
stable L\'evy vector depends on this measure,
and therefore is a non parametric distribution. However, as discussed below,
there is at least a trivial exception: the case of isotropic stable L\'evy
vectors.

Corresponding to our previous remarks, a (classical) stable L\'evy vector
has the following (Fourier) cumulant generating function:

\begin{equation}\label{eq_ddim_classical_K}
K_{\underline{L}}(\underline{k})= \delta t [i
(\underline{k},\underline{\gamma})
-\int_{\underline{u} \in \partial B_1}
(i\underline{k},\underline{u})^{\alpha} d\Sigma(\underline{u}) ] + o (\delta t)
\end{equation}

\noindent which yields with the help of the Eq.\ref{eq_ddim_fc_dK}:

\begin{equation}\label{eq_ddim_classical_Ktilda}
\frac{\partial }{\partial t} {\widetilde{K}}_{\underline{X}}(\underline{k})=
-div (\underline{m} + \underline{\underline{\sigma}}.\underline{\gamma})
- F^{-1}[\int_{\underline{u} \in \partial B_1}
(i \underline{\underline{\sigma
}}^{*}(\underline{x},t).\underline{k},\underline{u})^{\alpha}
d\Sigma(\underline{u}) ]
\end{equation}

The scalar case (Eq.\ref{eq_fc_dL}) corresponds to:

\begin{equation} \label{eq_1dim_dSigma}
 0 \le p \le 1:\beta= 2p-1,  ~ d\Sigma(u) = D cos ({{\pi \alpha} \over
{2}})[p \delta_{(u-1)} + (1-p) \delta_{(u+1)}]
\end{equation}

For any dimension d, the second term on the right hand side of
Eq.\ref{eq_ddim_classical_Ktilda}
corresponds to a fractional differentiation
operator of order $\alpha$.  This operator can be slightly re-arranged.
With the help of
the odd $d\Sigma^{-}(\underline{u})$ and even $d\Sigma^{+}(\underline{u})$
parts of the measure $d\Sigma(\underline{u})$,

\begin{equation}
2~d\Sigma^{+}(\underline{u}) = d\Sigma(\underline{u}) +
d\Sigma(-\underline{u});~~~
2~d\Sigma^{-}(\underline{u}) = d\Sigma(\underline{u}) - d\Sigma(-\underline{u})
\end{equation}

\noindent and the identity ($\theta$ being the Heaviside function):

\begin{equation} \label{eq_identity}
(ik)^{\alpha} =  |k|^{\alpha }
[\theta (k) e^{i{\alpha \pi \over  2}} + \theta (-k) e^{-i{\alpha  \pi\over
2}}]
\end{equation}

\noindent one can write the extension of Eq.\ref{eq_fokker_planck} under
the following form:

\begin{eqnarray}  \label{eq_ddim_fokker_planck}
{\frac{\partial }{\partial t}} p(\underline{x},t|\underline{x}_0,t_0)
 =  - div[ \underline{m}(\underline{x},t)
+\underline{\underline{\sigma }}(\underline{x},t).\underline{\gamma}) ]
p(\underline{x},t|\underline{x}_0,t_0) ]  \nonumber \\
 -   [ <(-\u{\u{\Delta}}:\u{\u{\sigma}}.\u{\u{\sigma}}^{*})^{\alpha \over
2}>_{\Sigma^{+}}
-<(\u{\nabla}.\u{\u{\sigma}}^{*}).(-\u{\u{\Delta}}:\u{\u{\sigma}}.\u{\u{\sigma}}
^{*})^{{\alpha - 1} \over 2}>_{\Sigma^-}]
p(\underline{x},t|\underline{x}_0,t_0)
\end{eqnarray}

\noindent where the symmetric fractional diffusive and
respectively the  antisymmetric advective-diffusive terms are defined ,
similarly
to Eq.\ref{eq_sigma_laplacian}, in the following manner:

\begin{eqnarray} \label{eq_frac.operators}
- <(-\u{\u{\Delta}}:\u{\u{\sigma}}.\u{\u{\sigma}}^{*})^{\alpha \over
2}>_{\Sigma^{+}} =
\int_{\underline{u} \in \partial B_1} d\Sigma^+(\underline{u})
F^{-1} [ \mid (\u{\u{\sigma }}^{*}(\u{x},t).\u{k},\u{u}) \mid^{\alpha}] \\
-
<(\u{\nabla}.\u{\u{\sigma}}^{*}).(-\u{\u{\Delta}}:\u{\u{\sigma}}.\u{\u{\sigma}}^
{*})^{{\alpha - 1}
\over 2}>_{\Sigma^-} =  \nonumber \\
\int_{\underline{u} \in \partial B_1} d\Sigma^-(\underline{u})
F^{-1} [ (- i \u{\u{\sigma }}^{*}(\u{x},t).\u{k},\u{u})\mid (\u{\u{\sigma
}}^{*}(\u{x},t).\u{k},\u{u})
\mid^{\alpha -1}]
\end{eqnarray}

In general, each term corresponds to a rather complex integration (which is
indicated
by the symbol $<.>_{\Sigma}$)by the measure $d\Sigma$ of directional
fractional Laplacians
(Eq.\ref{eq_direct_laplacian}).
However, the symmetric term becomes simpler as soon as the even part
$d\Sigma^+$  of the measure $d\Sigma$
is isotropic. Indeed, the integration over directions yields only a
prefactor $D$:

\begin{eqnarray} \label{eq_sym.frac.operator}
<-(\u{\u{\Delta}}:\u{\u{\sigma}}.\u{\u{\sigma}}^{*})^{\alpha \over
2}>_{\Sigma^{+}} =
D~ (-\u{\u{\Delta}}:\u{\u{\sigma}}.\u{\u{\sigma}}^{*})^{\alpha \over 2}
\nonumber \\
D = \int_{\underline{u} \in \partial B_1} d\Sigma^{+}(\underline{u}) \mid
(\u{u}_{1},\u{u}) \mid^{\alpha}
\end{eqnarray}

\noindent and for $\alpha =2$ this corresponds to the classical term
($\u{\u{\Delta}}:\u{\u{\sigma}}.\u{\u{\sigma}}^{*}$)
of the standard d-dimensional Fokker-Planck equation.  If $d\Sigma$ itself
is rotation invariant,
then the asymmetric operator vanishes, since $d\Sigma^- = 0$. If
furthermore, $\u{\u{\sigma}}$
is scalar, i.e. $\u{\u{\sigma}}= \sigma \u{\u{1}}$,
then one obtains the following Fractional Fokker-Planck equation:

\begin{eqnarray}  \label{eq_ddim_sym_fokker_planck}
{\frac{\partial }{\partial t}} p(\underline{x},t|\underline{x}_0,t_0)
&=& - div[ \underline{\underline{\sigma }} . \underline{\gamma}
(\underline{x},t) + \underline{m}(\underline{x},t)]
p(\underline{x},t|\underline{x}_0,t_0)  \\
&-& D~[(-\Delta )^{\alpha /2}] |\sigma (x,t)|^{\alpha}
p(\underline{x},t|\underline{x}_0,t_0)
\end{eqnarray}

Therefore, as one might expect it,the rotation symmetries yields a rather
trivial extension
of the standard Gaussian case: a fractional power $\alpha$ of the
d-dimensional Laplacian,
as in the pure scalar case (Eq.\ref{eq_fokker_planck}).  Obviously, the
integration performed
in Eq.\ref{eq_ddim_fokker_planck} is also greatly
simplified as soon as $d\Sigma(\underline{u})$ is discrete, i.e. its support
corresponds to a discrete set of directions $\u{u}_i$.

On the other hand, let us note that the framework of generalized stable
L\'evy vectors \cite{Generalized_levy_vector,Siegel,Meerschaert2,Jurek},
allows one to introduce
a much stronger anisotropy than the the measure $d\Sigma$ does it for
classical stable L\'evy vectors. This therefore diminishes the importance of
the asymmetry of the latter.
Indeed, the components of a generalized stable L\'evy vector do not
necessarily have the same L\'evy stability index, the latter being
generalized into
a second rank tensor. Similarly, the differential operators involved in the
corresponding
Fractional Fokker-Planck equation no longer have a unique order of
differentiation.
This is rather easy to check in case of a discrete measure
$d\Sigma(\underline{u})$
and we will explore the general case elsewhere.

\section{Existence and uniqueness of the solution}\label{sect.existence}
The previous sections established a generalization of the Fokker-Planck
equation for the evolution of the probability distribution of nonlinear
stochastic
differential equations driven by Levy stable noises. This was
the main goal of this paper. Naturally, one would also like to have if
possible a theorem of existence and uniqueness of the solution of
this equation. Due to its origin, such a theorem will
also imply that the solution will remain positive and
normalized, as required for a transition probability.
In this section we argue that the general results obtained \cite{Arnold} in
the
classical Gaussian case ($\alpha = 2$) are also relevant for the L\'evy
extension, whereas up until now existence and uniqueness conditions of partial
fractional differential equations have been scarcely explored (see
however Ref.\cite{Komatsu}, Henry) and therefore we cannot rely on general
results.

The classical Fokker-Planck equation  belongs
to the well explored domain of parabolic equations. Existence and
uniqueness of the solution fundamentally result \cite{Pazy} from the fact
that the
linear operator $A= -\Delta$ is a (self-adjoint) positive generator of a
semigroup
of contraction operators $T(t)= e^{-tA}, \: t \geq 0$. In the case
of constant coefficients (linear Langevin equation), the solution
is directly obtained with the help of $T(t)$ and this ensures its
existence, uniqueness and positiveness. Note that in our case, the
semigroup action corresponds to the equation of convolution (Eq.\ref
{eq.convolution}).

Similar results hold for a
Lipschitz variation of the coefficients, i.e.:

 \begin{equation}\label{eq.Lipschitz}
|m(x,t)-m(y,t)| + |\sigma (x,t) - \sigma (y,t)| \leq D| x- y |
  \end{equation}

  \noindent as well as a condition of slow growth in time of the coefficients
  $m(X(t),t)$ and $\sigma (X(t),t)$,e.g.:

  \begin{equation}\label{eq.slow.growth}
|m(x,t)| + |\sigma (x,t)| \leq  C|1+x|
  \end{equation}

\noindent  where $D$ and $C$ are given positive constants.

These conditions have been extensively used for the classical Fokker-Planck
equation
with non constant coefficients (e.g. \cite{Gardiner}).
Considering now the fractional generalization,
it is important to
note that the fractional power of the
Laplacian $-(-\Delta )^{\alpha /2} $ remains positive,
since its definition Eq.\ref{eq_riesz} corresponds to replacing the
eigenvalues $k^{2}$ by eigenvalues having as real part $|k|^{\alpha}$.
Therefore, we remain inside of the previous framework of contraction
semigroup and the previous results should hold.

This could be also seen on the integral form of the differential
equation. Indeed, in the classical case,  the Lipschitz condition is
classical for the Brownian forcing \cite{Oksendal, Janicki},
as well as for the more general case of martingale and
semi-martingale forcing \cite{Metivier, Kopp, Potter}. The latest case
is relevant for the stable L\'evy forcing. The Lipschitz condition can
be rather understood as a condition of convergence of the Picard
iteration method towards a fixed point:

 \begin{equation}\label{eq.Picard}
X^{n+1}(t) = X(t_{0}) +\int m(X^{n}(t),t) dt + \int \sigma(X^{n}(t),t)~
dL; X^{0}(t)= X(t_{0})
  \end{equation}

On the other hand, the condition of slow growth (\ref{eq.slow.growth}) in
time  prevents
a finite explosion time, i.e.
$X(t)$ remains finite for any given finite time $t$:
this condition is rather general, since it is already required by the
deterministic
part of the Langevin-like equation.

\section{Conclusion}\label{Conclusion}

We have derived a Fractional Fokker-Planck equation, i.e. a kinetic equation
which involves fractional derivatives, for the evolution of the
probability distribution of nonlinear stochastic differential equations
driven by non-Gaussian Levy stable noises. We first established this equation
in the scalar case, where it has a rather compact expression
with the help of fractional powers of the Laplacian, and then
discussed  and presented its extension to the vector case.
This  Fractional Fokker-Planck equation generalizes
broadly previous results obtained for a linear Langevin-like equation with
a L\'evy forcing, as well as the standard Fokker-Planck equation for
a nonlinear Langevin equation with a Gaussian forcing.

\section{Acknowledgments}\label{Acknowledgments}

We would like to thank L. Arnold, J. Brannan, D. Benson,
D. del-Castillo Negrete, D. Holm, P. Imkeller,
M. Meerschaert, I. Tchiguirinskaia, J. S. Urbach and W.A. Woyczynski
for helpful and stimulating discussions.

Part of this work was performed while
D. Schertzer was visiting Clemson University. We
acknowledge stimulating comments and
suggestions of an anonymous referee, which
helped us to sharpen many questions discussed in this
paper.


\end{document}